\documentstyle[12pt]{article}
\begin{document}
\def\a{\alpha}
\def\e{\varepsilon}
\def\be{\begin{equation}}
\def\ee{\end{equation}}
\def\l{\label}
\def\0{\setcounter{equation}{0}}
\def\T{\hat{T}_}
\def\b{\beta}
\def\S{\Sigma}
\def\3{d^3{\rm \bf x}}
\def\4{d^4}
\def\c{\cite}
\def\r{\ref}
\def\ba{\begin{eqnarray}}
\def\ea{\end{eqnarray}}
\def\n{\nonumber}
\def\R{\right}
\def\L{\left}
\def\q{\hat{Q}_0}
\def\X{\Xi}
\def\x{\xi}
\def\la{\lambda}
\def\s{\sigma}
\def\f{\frac}
\def\vx{{\rm \bf x}}
\def\j{\frac{\delta}{i \delta j_a ({\rm \bf x},x_0+t+t_1)}}
\begin{titlepage}
\begin{flushright}
      {\normalsize NI 94041\\
      SUSX-TH-95/6-2 }
\end{flushright}
\vskip 3cm
\begin{center}
{\Large\bf Real-time quantum field theory at finite temperature in an
inhomogeneous media}
\vskip 1cm

\mbox{Tengiz M. Bibilashvili}\footnote{On leave of absence from the
Institute of Physics,
Georgian Academy of Sciences, Tamarashvili str. 6,
Tbilisi 380077, Republic of Georgia,
e-mail:~tbib@physics.iberiapac.ge} \\
{
\it I.~Newton Institute for Mathematical Sciences,\\
 Cambridge CB3 0EH, England,\\
 School of Mathematical and Physical Sciences, University of Sussex,\\
 Brighton BN1 9QH, England}

\end{center}
\date{FEBRUARY 1995}
\vskip 1.5cm
\begin{abstract}
\footnotesize
The method of the real time perturbative calculations of
nonequilibrium averages is generalised to the case of varying chemical
potential.  Calculations are performed in the frame of Zubarev's nonequilibrium
density matrix approach.  In this approach perturbations of temperature and
other thermodynamical parameters are taken into account explicitly including
nonlinear terms.  It differs from the Schwinger-Keldysh approach through the
choice of more general initial conditions for the density matrix.
\end{abstract}

\end{titlepage}

\section{Introduction}
\setcounter{equation}{0}

Quantum field theory (QFT) at finite temperature in the equilibrium
\cite{Ma,M,DJ,NS,UMT} and nonequilibrium \cite{CH,RJ,U,Pa} cases has many
applications in study of the early Universe \cite{KSW,TK,HB} and heavy ion
collisions \cite{JK}.  Both kind of systems are intrinsically nonequilibrium in
nature, and only some details of their behaviour may be considered in the frame
of equilibrium description.

The purpose of this paper is to introduce a Green function generating
functional
for the nonequilibrium QFT at finite temperature and
chemical potential in the case when these thermodynamical parameters depend on
space and time coordinates.

The approach is based on Zubarev's nonequilibrium density matrix \c{Zub1}
(called by author nonequilibrium statistical operator (NSO) \c{Zub}).  The NSO
is a solution of the Liouville equation with initial conditions in the form of
a
local-equilibrium density matrix. This  matrix does not describe irreversible
processes such as thermal conductivity.  It corresponds to the maximum entropy
of a system for a given distribution of energy, charge density etc., but it is
not a solution of the Liouville equation, that rules the system dynamics.  This
deficiency plays no role if the initial condition in the form of a
local-equilibrium density matrix is fixed at the time $t \rightarrow - \infty$.
Hydrodynamical equations which account for the dissipative terms are derived
from the NSO explicitly (see \c{Zub}).

A situation with varying temperature is already analized in \c{B} and account
of
a chemical potential will be performed here in the same manner.  However, there
are some technical differences.  It is shown in \c{B} that consideration of
varying temperature requires the introduction of a more complicate time
integration contour in the complex time integration plane in comparison with
the
equilibrium case \c{NS}.  It will be shown here that in the case of a uniform
temperature distribution and varying chemical potential a complicated contour
is
not required.  Thus, temperature plays a special role in the quantum field
theoretical description.

There are other approaches to the study of QFT in inhomogeneous media.  It was
argued in \c{BS}, Zubarev's \c{Zub1,Zub} and McLeennan's \c{Mp,Mb} approaches
are the most useful, since they take into account thermal and charge
inhomogeneities explicitly.  These approaches have the same numerical results.

The most used method in QFT, that due to Schwinger-Keldysh \c{SK} is also based
on the Liouville equation but with equilibrium initial condition for the
density
matrix.  Only those thermal perturbations that occur after strong external
action on the initially equilibrium system are taken into account by this
approach .  It means that an inhomogeneous distribution of the initial values
of
thermal parameters is not taken into consideration.  And in general, this
approach does not determine an explicit expression for the temperature chemical
potential and other thermodynamical parameters.

\section{Density matrix}
\setcounter{equation}{0}

It is possible to consider nonequilibrium stationary systems in the frame of a
generalised Gibbs method.  The idea of local equilibrium is very useful for
this
aim.  It is based on the relaxation time property of nonequilibrium systems.
With this property, the system rapidly achieves a local equilibrium state and
then slowly relaxes to equilibrium \c{K}.  Investigation of the second long
stage is possible in terms of NSO.  As it is shown in \c{Zub}, the following
local equilibrium density matrix corresponds to the maximum of the system
entropy
\be
\hat{\rho}_l(t)=\X_l^{-1}e^{- \int\3\b^{\nu}(\vx,t)\T{0\nu}(\vx,t)+
\int\3\a(\vx,t)\hat{J}_0(\vx,t)},
\l{l}
\ee
where
\be
\b^{\nu}(x)=\b (x) u^{\nu} (x), \hspace{1.5cm} \a (x)=\b (x) \mu (x),
\l{lt}
\ee
$\b (x)=1/T (x)$ --- inverse local temperature, $u^{\nu}(x)$ ---
hydrodynamical velocity, $\mu (x)$ --- chemical potential corresponding
to the charge density $\hat{J}_0$
($\mu^{k}$ and $J_{0}^{k}$ if there are some conserved charges in the system),
$\T{\mu \nu}$ is the energy momentum tensor and $\X_l$ --- the partition
function that normalises (\r{l}).  It is clear that in the limit of uniform
values of temperature and chemical potential distribution we recover the
well-known grand canonical Gibbs density matrix, that in the rest frame (where
$u^{\mu}=(1,0,0,0)$) is
\be
\hat{\rho}_0=\X_0^{-1}e^{-\b\hat{H}+\a\q},
\l{e}
\ee
where $\hat{H}$ is the Hamiltonian of the system and
\be
\q=\int\3\hat{J}_0(\vx,t)
\l{Q}
\ee
its charge.

The density matrix (\r{l}) gives the correct results for the energy and charge
density.  But it doesn't describe dissipative processes such as thermal
conductivity, diffusion, viscosity etc.

A density matrix that completely describes the dynamics of the system including
dissipative processes is obtained by Zubarev as a retarded solution of the
Liouville equation \c{Zub}
\begin{equation}
\frac{\partial \hat{\rho}}{\partial t}-i[\hat{\rho},\hat{H}]=0
\label{le}
\end{equation}
with the initial condition in the form of a local-equilibrium density matrix
(\r{l}).  For this aim an external source that breaks time invariance of the
equation is introduced in the right hand side of (\r{le})
\begin{equation}
\frac{\partial \hat{\rho}_{\e}}{\partial t}-i[\hat{\rho_{\e}},\hat{H}]=
-\e (\hat{\rho}_{\e}-\hat{\rho}_{l}).
\label{lee}
\end{equation}
The limit $\e \rightarrow 0$ is to be taken after thermodynamical averaging.
The same procedure of the retarded solution selection of the equation of
motion with the free initial conditions is used in scattering theory
\c{GG}. The solution of (\r{lee}) is \c{Zub1,Zub}
\be
\hat{\rho}_{\e}(t)=\e\int_{-\infty}^{0}dt_1
e^{\e t_1}\hat{\rho}_{l}(t+t_1).
\l{nso}
\ee
Thus, integration of (\r{lee}) with the positive sign of $\e$ gives the
retarded solution of the Liouville equation (\r{le}). This solution
corresponds to the entropy
\be
S=-{\rm Tr}\{\hat{\rho_{\e}}\ln\hat{\rho_{\e}}\}
\l{ent}
\ee
increase.  There are some other useful techniques to obtain the smooth density
matrix describing entropy increase (see for example \c{Zub}).
The exact solutions $\hat{\rho}_{exact}$ of
(\r{le}) corresponds to the conserved entropy $S=-{\rm Tr} \{\hat{\rho}_{exact}
\ln\hat{\rho}_{exact}\}$.  It may be averaged in quantum states with close
energies or in time intervals to obtain coarse grained
\be
\hat{\rho}_{\Gamma}(t)=\frac{1}{\Delta\Gamma}{\rm Tr}_{\Delta \Gamma}
\hat{\rho}_{exact}(t)
\l{cg}
\ee
or time averaged
\be
\hat{\rho}_{\tau}(t)=\frac{1}{T}\int_{t}^{t+\tau}d t' \hat{\rho}_{exact}(t')
\l{ta}
\ee
density matrices.  So, $S_{\Gamma}=-{\rm Tr} \{\hat{\rho}_{\Gamma}
\ln\hat{\rho}_{\Gamma}\}$ and $S_{\tau}=-{\rm
Tr}\{\hat{\rho}_{\tau}\ln\hat{\rho}_{\tau}\}$ increase in time.

In the general case, matrix (\r{l}) is expressed \c{ZPS,SS}
\be
\hat{\rho}_l=\X_l^{-1}e^{- \int d\s^{\mu}(\b^{\nu}(\vx,t)\T{\mu \nu}(\vx,t)-
\a (\vx,t)\hat{J}_{\mu}(\vx,t))},
\l{lg}
\ee
where $\s_{\mu}$ defines some space-like hypersurface.
Now it is possible to introduce transformed energy-momentum tensor depending
on chemical potential.
\ba
\hat{\rho}_l=\X_l^{-1}e^{- \int d\s^{\mu}(\b^{\nu}(\vx,t)\T{\mu \nu}(\vx,t)-
\mu (\vx,t)\b^{\nu}(\vx,t)u_{\nu}(\vx,t)\hat{J}_{\mu}(\vx,t))}=
\n \\
=\X_l^{-1}e^{- \int d\s^{\mu}\b^{\nu}(\vx,t)\T{\mu \nu}'(\vx,t)},
\l{lg1}
\ea
where
\be
\T{\mu \nu}'(x)=\T{\mu \nu}(x)-\mu (x)u_{\nu}(x)\hat{J}_{\mu}(x),
\l{tst}
\ee
and identification
\be
\b (x)\equiv \b^{\nu}(x)u_{\nu}(x)
\ee
was used.

When the distance of the hydrodynamical parameter variations $L$ is bigger
than the correlation lengths
\be
L\gg l_{corr}
\ee
the hypersurface $\s^{\mu}$ may be chosen in an arbitrary way \c{ZPS} and it
is possible to consider the hypersurface for which $dx^{0}=0$ ($d\s^{\mu}=
(\3,0,0,0)$). In
this case NSO may be written in the form (\r{nso})
\be
\hat{\rho}_{\e}=
\e\int_{-\infty}^{0}dt_1 \Xi^{-1}_le^{\int \3
\b^{\nu}(\vx,t+t_1)\T{0\nu}'(\vx,t+t_1)},
\l{NSOS}
\ee
with $\T{0\nu}'$ defined by (\r{tst}).

\section{Thermal perturbations}
\setcounter{equation}{0}

Consider the inverse temperature distribution slightly differing from a
certain constant value $\b^{\mu}_0=\b_0u^{\mu}_0$ by a small parameter
$\b_1^{\mu}(x)$
\be
\b^{\mu}(x)=\b_0^{\mu}+\b_1^{\mu}(x).
\l{+}
\ee

Using this expression it is possible to apply our results \c{BP,B,BS} but with
the transformed energy-momentum tensor (\r{tst}) in the NSO (\r{NSOS}).
NSO may be written in the rest frame ($u^{\mu}_0=(1,0,0,0)$). In this frame
$\b^{\mu}_{0}\hat{P}_{\mu}=\b_0 \hat{H}$
\begin{eqnarray}
\hat{\rho}_{\varepsilon}(t)= \varepsilon\int_{-\infty}^{0}
d t_1 e^{\varepsilon t_1}
\Xi_{l}^{-1}\exp \left[-\beta _0 \hat{H}'- \right.
\; \; \;  \nonumber \\ \left.
- \int \3 \beta^{\mu}_1 ({\bf x},t+t_1)
\hat{T}_{0 \mu}'({\bf x},t+t_1) \right]
\label{9}
\end{eqnarray}
where
\be
\hat{H}'=\hat{H}-\int \3 \mu(x)\hat{J}^0(x).
\l{sh}
\ee

Now we make use of a well-known formula for the exponent of noncommuting
operator sums \c{FKS} rewritten for our case in \c{BP},
\begin{equation}
\exp\left[- \beta_0 \hat{H}'+\hat{B}(\tau)\right]=
e^{- \beta_0 \hat{H}'} {\rm T}_C \left\{\exp \left[i \int_{C} d x_0
\beta_{0}^{-1}\hat{B}(\tau+ x_0) \right] \right\},
\label{11}
\end{equation}
where $C$ is an arbitrary contour in the complex time-plane appropriate for the
real time QFT \c{M,NS,LvW,RR,HK} and
${\rm T}_C$ --- is an ordering by the time variable along $C$.
With the help of (\r{11}), NSO may be presented as
\begin{eqnarray}
\hat{\rho}_{\varepsilon}(t)=\varepsilon \int_{-\infty}^{0}
d t_1 e^{\varepsilon t_1} \X_{l}^{-1}
e^{-\beta _0 \hat{H}'} \times \; \; \; \; \nonumber \\
\times   {\rm T}_C \left\{\exp \left[
-i \int_{C} d^4 x \frac{\beta^{\mu}_1 ({\bf x},t+t_1)}{\beta_0}
\hat{T}_{0 \mu}'({\bf x},x_0+t+t_1) \right] \right\}.
\label{11b}
\end{eqnarray}

Here the chemical potential is to be considered.  After a reduction (\r{sh}),
term with the chemical potential became part of the Hamiltonian $\hat{H}'$. In
analogy with (\r{+}), it may be presented as
\be
\mu(x)=\mu_0+\mu_1(x)
\l{+cp}
\ee
and the part of $H'$ dependent on $\mu_1$ will be considered as a perturbation.

Perturbations of thermodynamical parameters are called ``thermal
perturbations''
\c{K} differing from ``mechanical'' ones which may be considered as part
of the Hamiltonian. However after reduction (\r{sh}) of the Hamiltonian,
perturbations of the chemical potential may be formally taken into account as
mechanical ones. Analogous technical tricks are not possible with the
temperature
and it leads to the more complicated contour for the time integration \c{B,BS}.

Thus, instead of an ordinary perturbation in the Hamiltonian $V$ (usually
characterised by the coupling constant) we have a more complicated one
\be
\hat{V}\rightarrow \hat{V}-\int \3 \mu_1(x)\hat{J}^0(x)
\l{pert}
\ee
dependent on both coupling constant and chemical potential deviation
$\mu_1(x)$.
The first term in (\r{pert}) corresponds to the familiar vertex and the second
one will be presented in the diagram as an insertion into the line.

\section{Generating functional of the theory}
\setcounter{equation}{0}

Now a generating functional for the Green's functions may be considered
\be
{\cal Z}[j]={\rm Tr}\left\{\hat{\rho}_{\varepsilon}{\rm
T}e^{j\hat{\phi}}\right\}=
\varepsilon \int_{-\infty}^{0}
d t_1 e^{\varepsilon t_1} N(t_1){\rm Tr}
\left\{
\hat{\rho}_{0}
{\rm T}_C \left[e^{\b_1\hat{T}'}\right]
{\rm T} \left[e^{j\phi}\right]
\right\},
\l{qq}
\ee
where
\begin{equation}
e^{j\hat{\phi}}=\exp \left[i\int d^4yj(y)\hat{\phi}(y)\right].
\label{A3}
\end{equation}
\be
e^{\b_1\hat{T}'}=\exp \left[
-i \int_{C} d^4 x \frac{\beta^{\mu}_1 ({\bf x},t+t_1)}{\beta_0}
\hat{T}_{0 \mu}'({\bf x},x_0+t+t_1) \right]
\l{}
\ee
and $N(t_1)$ is the residual part from the partition function $\Xi_l$.

Using the identification
\begin{equation}
F(\hat{\phi})e^{j\hat{\phi}}\equiv
F\left[\frac{\delta}{i\delta j}\right]e^{j\hat{\phi}}
\label{id}
\end{equation}
we may rewrite (\r{qq}) in the form
\begin{eqnarray}
{\cal Z}[j]=
\varepsilon \int_{-\infty}^{0} d t_1 e^{\varepsilon t_1}
\times \nonumber \\ \times \left.
\exp \left[-i \int_{C} d^4 x \frac{\beta^{\mu}_1 ({\bf x},t+t_1)}{\beta_0}
T_{0 \mu}'\left[\frac{\delta}{i\delta j({\bf x},x_0+t+t_1)}\right]
\right] Z[j,j'] \right|_{j'=0},
\label{G3}
\end{eqnarray}
where $T_{\mu \nu}'[\delta/i\delta j]$ is the same function of
$\delta/i\delta j$ as $T_{\mu \nu}'(\phi)$ of $\phi$ and
\be
Z[j,j']={\rm Tr}\left\{\hat{\rho}_0{\rm T}_C[e^{j\hat{\phi}}]
{\rm T}[e^{j'\hat{\phi}'}]\right\}
\l{GF4}
\ee
The generating functional (\r{GF4}) may be defined with the unit time-ordering
procedure if the time-integration contour ${\cal C}$ (fig.  1) is introduced.
This contour differs from the one considered in the previous papers \c{B,BS},
however it satisfies all the conditions used there.  Segment ${\cal C}_1$
is one where external fields fixed in the second
exponent in ({\r{GF4}) and ${\cal C}_{2''}$ and
${\cal C}_3$ --- for thermal perturbations in the first time-ordered exponent
in
the expression (\r{GF4}). As previously described, the vertical
part is considered at $-\infty$ and, therefore, it plays no role.  A difference
between the contour indices $2'$ and $2''$ will be observable from the region
of
time integration in the generating functional and it is not specified by the
separate index value of $\phi_a$ and $J_a$.  This contour consists only of 3
horizontal segments and the theory is based on the $3\times 3$ free propagators
(instead $4\times 4$ used previously \c{B,BS}).

It is clear that the generating functional has the same form as in \c{B}:
\ba
{\cal Z}[j_1,j_2,j_3]=\e \int_{-\infty}^{0}dt_1 e^{\e t_1}
\exp \left\{-i\int d^4 x \sum_{a=1}^3(-1)^aV\left[\j\right]\right\}
\times \n \\ \times
\exp \left\{i\int_{-\infty}^{0}dx_0\int \3
\frac{\b _1^{\mu}(\vx ,t+t_1)}{\b _0}\sum_{a=2}^3(-1)^a T_{0\mu}'
\left[\j \right]\right\}\times \n \\ \times
\exp \left\{-i\int d^4x\sum_{a=1}^{3}(-1)^a\mu _1(\vx,t+t_1)J_0
\left[ \j \right]\right\} \times \l{e11}\\ \times
\exp \left\{-\frac{i}{2}\int d^4z_1d^4z_2\sum_{a,b=1}^{3}
j_a(z_1){\cal D}_{\b _0 \mu _0}^{ab}(z_1-z_2)j_b(z_2)\right\}.
\n
\ea
The difference with the generating functional derived in \c{B} consists in the
presence of the chemical potential perturbations $\mu_1$ and in the free Green
function that depends not only on the constant part of the temperature $\b _0$,
but also on the constant part of the chemical potential $\mu_0$
Free propagators dependent on the chemical potential are well-known in the
equilibrium case (see for example  \c{Kow,LvW})

Finally I would like to mention that in the case of a constant temperature and
varying chemical potential term, $\b_1$ is absent in (\r{11b}) and no
double-time ordered product occur in (\r{qq}).  It makes it possible to
consider
a theory in this particular case with the ordinary contour that is suitable in
the equilibrium case, where free propagators are $2\times 2$ matrices.

It is important to mention also that this theory is free from pinch
singularities. Arguments for this fact are the same as in \c{B,BS}.

\vspace{0.2in}
{\Large \bf Acknowledgements}
\vspace{0.2in}

I am thankful to N.~Dombey, T.~Kibble
and J.~Wright for their hospitality at The University of Sussex and I.Newton
Institute at The Cambridge University. I would like to acknowledge
P.~Landshoff, R.~Rivers and J.C.~Taylor for useful remarks. Special
thanks to E.~Copeland for detailed discussions of this text.
This visit was supported
financially by The Royal Society.

\newpage
\begin{figure}[ht]
\begin{picture}(450,210)
\put(60,190){\makebox(0,0){$-\infty$}}
\put(33,150){\makebox(0,0){$-\infty -2i\varepsilon$}}
\put(60,25){\makebox(0,0){$-\infty-i\beta_0+i\varepsilon$}}
\put(350,150){\makebox(0,0){$+\infty-i\varepsilon$}}
\put(225,210){\makebox(0,0){\em Im\ t}}
\put(360,190){\makebox(0,0){\em Re\ t}}
\put(160,190){\makebox(0,0){\em $t_1$}}
\put(160,24){\makebox(0,0){\em $t_1-i\beta _0$}}
\put(330,175){\makebox(0,0){$_{{\cal C}_1}$}}
\put(260,149){\makebox(0,0){$_{{\cal C}_{2'}}$}}
\put(120,162){\makebox(0,0){$_{{\cal C}_{2''}}$}}
\put(120,45){\makebox(0,0){$_{{\cal C}_{3}}$}}
\put(70,90){\makebox(0,0){$_{{\cal C}_4}$}}
\put(150,120){\makebox(0,0){${\cal C}$}}
\put(15,180){\line(1,0){355}}
\put(210,15){\line(0,1){195}}
\multiput(60,150)(2,0.1){150}{.}
\multiput(360,165)(-2,0.15){100}{.}
\multiput(60,37,5)(0,2){57}{.}
\multiput(160,30)(-2,0.15){50}{.}
\end{picture}
\caption{Contour for the nonequilibrium case}
\end{figure}
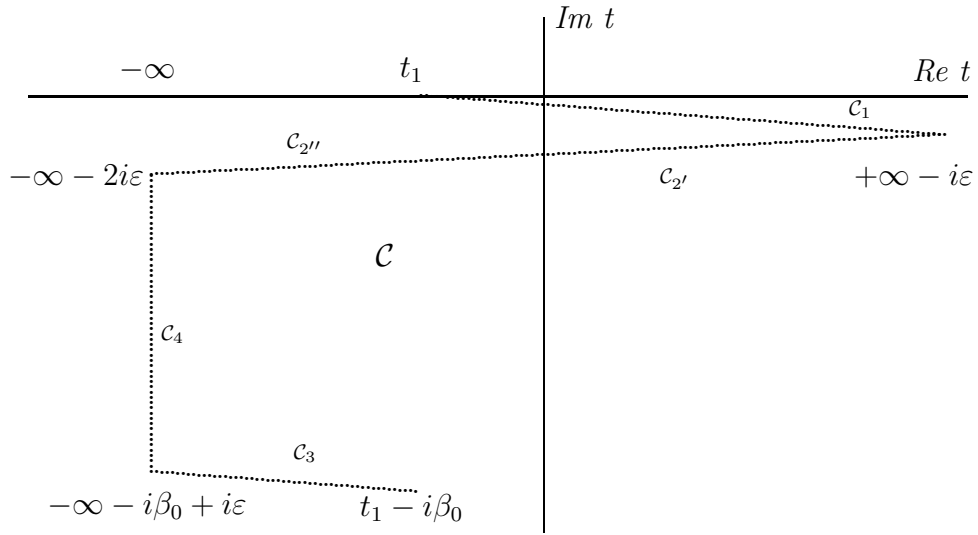

\end{document}